\pdfoutput=1 % if your are submitting a pdflatex (i.e. if you have
\documentclass[twocolumn,epjc3]{svjour3}
\usepackage{siunitx}
\usepackage{todonotes}
\usepackage{subfig}
\usepackage[export]{adjustbox}
\usepackage{placeins}
\usepackage{amsmath}
\usepackage{mathtools, cuted}

\usepackage{amsmath, amssymb, mathtools} % Математика
\usepackage{graphicx} % Для изображений
\usepackage{siunitx} % Для единиц измерения
\usepackage[export]{adjustbox} % Доп. настройки графики

\usepackage{hyperref} % ВСЕГДА подключать последним!
\hypersetup{
    unicode=true,
    psdextra=true,
    colorlinks=true,
    citecolor=blue,
    linkcolor=red
}

\usepackage[14pt]{extsizes} % Правильно меняет размер текcта
\usepackage[utf8]{inputenc}	%Кодировка utf8	
\usepackage[T2A]{fontenc} % поддержка кириллицы
\usepackage{pdfpages}
\usepackage{mathabx}
\usepackage{enumitem}
\begin{document}
\title{Intrinsic Resolution of Organic Scintillators}

\author{O.\ Smirnov}

%\authorrunning{Short form of author list} % if too long for running head

\institute{Joint  Institute  for  Nuclear  Research,  141980,  Dubna,  Russian Federation
              \email{osmirnov@jinr.ru}        %  \\
%             \emph{Present address:} of F. Author  %  if needed
%           \and
%INR
}

%\author[a]{O.\ Smirnov }

%\affiliation[a]{Joint  Institute  for  Nuclear  Research,  141980,  Dubna,  Russian Federation}

% e-mail addresses: only for the corresponding author
%\email{osmirnov@jinr.ru}

\maketitle
\begin{abstract}
Fluctuations in photon production within scintillators contribute to the overall energy resolution of scintillation detectors. While typically negligible, this contribution - known as intrinsic resolution (IR) - presents significant challenges for both experimental measurement and Monte Carlo simulation. Using the phenomenological model proposed in~\cite{MyIntrinsic}, we interpreted multiple datasets of IR measurements in organic scintillators. Although my results generally agree with the IR dataset examined in~\cite{MyIntrinsic}, a comprehensive evaluation of systematic uncertainties highlights the necessity for more detailed investigation. Based on my findings, we offer specific recommendations for future measurements.
\end{abstract}

\section{Introduction}

Intrinsic Resolution (IR) of a scintillator at a given energy represents the fundamental limit of energy resolution achievable when an ideal, spatially uniform detector with 100\% efficiency detects all scintillation photons. Under normal fluctuations in photon production, this theoretical limit is determined  solely by statistical fluctuations, resulting in an energy resolution of $\frac{1}{\sqrt{\overline{N_\text{ph}}}}$, where $\overline{N_\text{ph}}$ is the mean number of detected photons.

Deviations from normal photon statistics can significantly impact the sensitivity of high-resolution spectroscopy experiments. Accurate quantification of these effects is therefore essential. This is particularly critical for experiments requiring exceptional energy resolution, including neutrinoless double beta decay searches and next-generation neutrino detectors like JUNO~\cite{JUNO}. While multiple factors influence the practical energy resolution of these detectors, IR represents a potentially significant contributor.

In my previous work~\cite{MyIntrinsic}, we proposed a simple single-parameter phenomenological model to describe the energy dependence of IR’s contribution to scintillation line width. The model assumes that the variance in photon production scales linearly with the mean photon yield, with its sole parameter representing the IR contribution at 1 MeV. Once calibrated using a setup with known photoelectron yield, this model allows for IR calculations across any detector configuration and energy within the model's range of validity.

In this paper, we analyse energy resolution data for various organic scintillators reported by multiple research groups~\cite{Swiderski:intrinsic:2012,Formozov:2019,Formozov:PhD,BC408ep,EJ200,Deng,BrxPhaseI,Pablo_PhD,BrxSpectroscopy}. The primary objective of this analysis is to assess the consistency between experimental measurements and the theoretical model. My results demonstrate that there are no statistically significant deviations from the model in the 10 keV to 4 MeV energy range. Although the model agrees well with the data, accurate parameter estimation is complicated by potential systematic uncertainties, particularly those relating to the calibration of the photomultiplier tubes (PMTs) used in the measurements.

Additionally, we investigate potential systematic uncertainties involved in extracting the contribution of intrinsic energy resolution from the total scintillation line width in the non-ideal scintillation detector, and provide recommendations for further studies.

The origin of the non-Gaussian deviations in photon production that are responsible for the IR lies beyond the scope of this paper.

\section{\label{StatDesc}Statistical description of the scintillations counter}

In this section we review the classical description of the statistical properties of scintillation counter, following the formalism in~\cite{Birks}, and introduces the framework for accounting for the contribution of the intrinsic resolution (IR). For simplicity, we restrict the analysis to point-like energy deposits.

In an ideal scintillation counter equipped with a single photomultiplier tube (PMT), light detection can be modelled as a three-stage cascade process:

\begin{description}[font=\bfseries, leftmargin=0.1cm]
    \item[photon emission:] the first stage involves the emission of $N_\text{ph}$ photons, characterised by mean photon yield  $\overline{N_\text{ph}}$ and relative variance $v(N_\text{ph})$;
    
    \item[photon/p.e. transfer:]  the transfer process includes several binomial processes: the (geometric) probability that a photon will reach the photocathode of the PMT, the probability that a photon will be converted into a photoelectron on the photocathode, and the probability that a photoelectron will survive on its way to the first dynode of the electron multiplier. Because of the binomial character of all the processes involved, the final distribution is also binomial, with the resulting transfer probability $p$;
    
    \item[p.e. multiplication:] the final stage involves statistical amplification of photoelectrons in the PMT’s electron multiplier. Let $G$ denote the average multiplication of a single electron and $v_{1}$ the relative variance of the anode signal corresponding to a single-electron response of the PMT.
\end{description}

In the general case, the average signal at the output of a three-stage cascade is~\cite{ShockleyPierce}:
  \begin{equation}
      \overline{X}=\overline{X_1}\cdot\overline{X_2}\cdot\overline{X_3}, \label{eq:Mean}
  \end{equation}
and its relative variance is
  \begin{equation}
      v(X)=v(X_1)+\frac{v(X_2)}{\overline{X_1}}+\frac{v(X_3)}{\overline{X_1} \cdot\overline{X_2}} \label{eq:Var}.
\end{equation}

Here $X_i$ is a random quantity at stage $i$; $\overline{X_i}$ and $v(X_i)$ are its mean and relative variance. The relative variance is defined as the ratio of the variance to the mean squared.

When applied to the scintillation counter, equation (\ref{eq:Mean}) takes the form~(see~\cite{Birks} and references therein):

  \begin{equation}
      \overline{Q_\text{A}}=\overline{N_\text{ph}}\cdot p \cdot\overline{G}=\overline{Q}\cdot\overline{G}, \label{eq:MeanIdeal}
  \end{equation}
where $Q_\text{A}$ is the total amount of electrons at the PMT's anode (total charge), and
$\overline{Q}=\overline{N_\text{ph}}\cdot p$ is the average number of photoelectrons corresponding to the detected average anode charge $\overline{Q_\text{A}}$.  
Giving that the distribution of the random transfer efficiency is binomial with relative variance $\frac{1-p}{p}$, equation~(\ref{eq:Var}) can be rewritten as follows:

\begin{multline}
v(Q_\text{A})=v(N_\text{ph})+\frac{1-p}{p}\frac{1}{\overline{N_\text{ph}}}+
\\
+\frac{v_1}{\overline{N_\text{ph}} \cdot p}=
\\
=\big{(}v(N_\text{ph})-\frac{1}{\overline{N_\text{ph}}}\big{)}+\frac{1+v_1}{Q}. 
\label{eq:VarIdeal}  
\end{multline}

It is convenient to measure the anode charge $Q_\text{A}$ directly in photoelectrons, given that the factor $\overline{G}$ does not enter into the expression for the relative variation of the anode charge. For this purpose, the PMT anode charge should be calibrated directly in photoelectrons. Note that the amount of photoelectrons $Q$ detected at the anode is not integer because of the detection process, and (\ref{eq:VarIdeal}) is the same for variables $Q$ and $Q_\text{A}$. In the following discussion, we will use photoelectrons count, $Q$, instead of the anode charge, $Q_\text{A}$.

For a non-ideal detector, the transfer probability $p$ may depend on multiple factors, including the photon's point of origin, the details of the photon transfer process from that point to the photocathode, the angle and point of incidence on the photocathode and the efficiency of photoelectrons collection on the first dynode. Each set of possible transfer factors, "i", is characterised by a binomial distribution with a fixed average transfer probability $p_i$. The average transfer probability, $\overline{p}$, and its variance, $v(p)$, should be calculated across all possible sets of factors. %Note that  $v(p)$ is the variance of average probabilities $p_i$ here, rather than the variance of the random variable, $p$, as in the ideal case.
Note that notation $v(p)$ refers to the variance of the values of average probabilities $p_i$, in contrast to the variance of the random binomial variable with a fixed average probability $p$  in the previously considered case.

The average p.e. count at the anode is $\overline{Q}=\overline{N_\text{ph}}\overline{p}$. The variance, $Var(Q)$, of the $Q$ distribution can be obtained by summing the averages of the squares of the individual contributions from each set of factors and then subtracting the mean value:
\begin{multline}
    Var(Q)=\sum_i w_i\overline {Q_i^2} -\overline{Q}^2=
    \\
=\sum_i w_i \big{(} \overline {Q_i}^2+(1+v_1)Q_i+
\\
+(v(N_\text{ph})-\frac{1}{\overline{N_\text{ph}}})\overline {Q_i}^2\big{)}-\overline{Q}^2.
\label{eq:VarQ_1}
\end{multline}

Here $Q_i\equiv \frac{p_i}{\overline{p}}  \overline{Q}$ and $w_i$ is the weight of the corresponding set of transfer factors. Equation~(\ref{eq:VarIdeal}) was used to develop $\overline{Q_i^2}$.
Noting that $\sum_i w_i Q_i^2=\overline{Q}^2 (1+v(p))$ and $\sum_i w_i Q_i=\overline{Q}$, we can rewrite (\ref{eq:VarQ_1}) as

\begin{multline}
Var(Q)=\overline{Q}(1+v_1)+v(p)\overline{Q}^2+
\\
+(v(N_\text{ph})-\frac{1}{\overline{N_\text{ph}}})(1+v(p))\overline {Q}^2,
\label{eq:VarQ_2}
\end{multline}

and the relative variance of the p.e. count is expressed as follows:

\begin{multline}
v(Q)\equiv \frac{Var(Q)}{\overline{Q}^2}=\frac{1+v_{1}}{\overline{Q}}+v(p)+
\\
+(1+v(p))\left[v(N_\text{ph})-\frac{1}{\overline{N_\text{ph}}}\right].
\label{Birks}
\end{multline}

The statistical fluctuations of the number of scintillation photons, $N_\text{ph}$, in a scintillation event (the last term in (\ref{Birks})) only give a non-zero contribution if the relative variance of photon production, $v(N_\text{ph})$, exceeds the relative variance of the normal distribution, $v(N_\text{ph})>\frac{1}{\overline{N_\text{ph}}}$. If this is the case and the last term is not negligible, formula~(\ref{Birks}) contains both photoelectron and photon counts, making it impractical.

In ~\cite{MyIntrinsic}, a phenomenological model was proposed for the contribution of the term responsible for the intrinsic resolution. This model characterises the excess of the relative variance $v(N_\text{ph})$ over the relative variance of the normal distribution. This excess is scaled as $\frac{Q_{1}}{Q}$, where, $Q_{1}$ corresponds to the total charge collected for an event with an energy deposit of 1~MeV :

\begin{equation}
    v_\text{int}\equiv\frac {Q_1}{Q} v_{1}^\text{int}.
    \label{eq:v_int}
\end{equation}

The universal parameter $v_{1}^\text{int}$ in (\ref{eq:v_int}) corresponds to the relative scintillation signal variance due to the intrinsic resolution at an energy deposit of 1 MeV. The multiplier $\frac{Q_{1}}{Q}$ accounts for the non-linearity of the energy conversion to the average number of detected photoelectrons. In the region of proportionality,
it is equivalent to $\frac{1}{E}$. The value of $Q_1$ depends on the properties of the scintillator and the characteristics of the detector.

The value of $v(p)$ is typically very small compared to unity, and the term $(1+v(p))$ can be omitted without introducing significant bias.

The only assumption in (\ref{eq:v_int}) is that the variance of the signal is proportional to the amount of detected photoelectrons, which requires independent verification. In this paper, we test this assumption using available sets of previously published data. Specifically, we assess the consistency of the relative variance $v(Q)$ of the PMT signal with the average count of detected photoelectrons (p.e.) $Q$, given by

\begin{equation}
v(Q)=\frac{1+v_{1}}{Q}+v(p)+(1+v(p))\frac {Q_1}{Q} v_{1}^\text{int},
\label{Birks2}
\end{equation}

with the available data. Henceforth,  we adopt the notation $Q$ instead of $\overline{Q}$ in formulae.

As can be seen from equation~(\ref{eq:v_int}), the additional variance due to the IR contributes in the same way as the statistical contribution, $\frac{1+v_{1}}{Q}$, as they have the same dependence on $Q$. Therefore, within the framework of the considered formalism, the IR effect imitates the statistical smearing of the charge signal. This effect is more pronounced for detectors with a high photoelectron yield. At an energy deposit of 1~MeV (i.e. at $Q=Q_1$), the IR contribution $v_1^\text{int}$ should be compared to the value $\frac{1+v_1^\text{Det}}{Q_1}$: the higher $Q_1$, the higher the relative contribution of the IR. 

In any study on energy resolution, the measured value is $v(Q)$. For the correct extraction of the IR component, which is a differential measurement, the experimentalist should provide a precise estimate of the statistical and non-uniformity terms, $v(p)$. The precision of the measurements of these quantities must significantly exceed the expected contribution of the IR.

Taking into account the low values of $v_1^\text{int}$, which are typically of the order of $10^{-4}$ (see \cite{MyIntrinsic} and table~\ref{Table:AllIRData}), it is clear that an experiment involving a high yield of photoelectrons would be more suitable for measuring $v_1^\text{int}$. For example, with an expected IR of 2\% at 1~MeV ($v_1^\text{int}=4\times 10^{-4}$) and a relative yield of 100 p.e./MeV, the IR contribution is $10^{2}\times 0.02^2=0.04$, which is small compared to $1+v_1$ for any $v_1$. For a relative yield of 500 p.e./MeV and the same  $v_1^\text{int}$, the IR contribution reaches 0.2, making it easier to measure. Another important factor in this type of measurement is the PMT resolution in a single p.e. mode, characterised  by the parameter $v_{1}$. As it contributes in the same way as the IR, any uncertainty in its definition translates into the uncertainty in the IR effect. Due to the complex structure of the PMT response and the presence of noise, measuring the $v_1$ itself is a complicated task.

Last but not least, the complication of the p.e. counting is the calibration of the PMT anode signal in p.e., i.e. the definition of the coefficient linking the experimentally measured values (e.g. the amplitude of the output signal or the integral of the charge, etc.) and the true amount of photoelectrons. Any bias in the definition of the calibration coefficient leads to the shift in the detected p.e scale. If the real charge is related to the measured one as $Q_\text{measured}=c\cdot Q_\text{real}$, the statistical contribution in (\ref{Birks2}) should be scaled, and the formula transforms to:

\begin{multline}
v(Q_\text{measured})=c\frac{1+v_{1}}{Q_\text{measured}}+v(p)+
\\
+(1+v(p))\frac {Q_1}{Q} v_{1}^\text{int}.
\label{Birks3}
\end{multline}

It can be seen that the statistical term in the observed relative variance depends on the calibration bias,  whereas the IR contribution is defined in a way excluding dependence on the PMT calibration.

In this paper, I will use the notation $Q$ for the measured charge, while keeping the calibration parameter "c".

I would like to highlight the implicit assumption of linearity in the conversion from $N_\text{ph}$ in $Q$, and the explicit one concerning the monoenergetic nature of the source. If the source is not monoenergetic, the term of the type $\frac{\Delta Q^2}{Q^2}$ should be taken into account, where  $\Delta{Q}^2$ corresponds to the source energy variance.  

Formula (\ref{Birks3}) still omits some possible contributions, such as the constant dark noise contribution or digitiser rounding. 

When studying energy resolution, we are always dealing with the experimentally observed quantity: the amount of p.e., $Q$.
For a monoenergetic electron source, the average amount of produced photoelectrons is related to the particle energy as follows:

\begin{equation}
    Q(E)=LY\cdot E \cdot f_\text{NL}(E), \label{eq:Q(E)}
\end{equation}

where $LY$ is the relative p.e. yield at 1 MeV, the electron energy, $E$, is expressed in MeV, and the function $f_\text{NL}(E)$ accounts for possible nonlinearities in the conversion of electron kinetic energy into light. Note that similar relations can be used for other particles ($\gamma$, protons, $\alpha$-particles, etc.) with their corresponding light yield ($LY$) and $f_\text{NL}(E)$ functions.

Since all the measurements considered in this paper are reported on an energy scale, we use the approximation $v(Q)\simeq v(E)$. This approximation is justified because the authors apparently used the inverse approximation, $v(E)\simeq v(Q)$, implicitly, to obtain the energy resolution. 

Also, the nonlinearity function and the light yield provided for each measurement were used to convert the energy back to p.e.. The last step would, of course, be unnecessary if the measurements were provided directly on the p.e. scale, without conversion to energy.

\section{The intrinsic resolution for gamma's}

In the actual analysis, we are not dealing with gamma rays, although the resolution for the fully absorbed gamma-line is also important in scintillation experiments. That is why we would like to emphasise the difference in intrinsic resolution between gamma's and electrons.
In the typical energy range of interest of up to a few MeV, $\gamma$'s lose their energy through a series of Compton scattering on electrons, and the total registered signal (amount of p.e.) is the sum of the signals from the individual Compton electrons:

\begin{equation}
    Q_{\gamma}=\sum_i Q_e(E_i),
    \label{eq:Q_gamma}
\end{equation}

where $Q_{\gamma}$ is the amount of p.e. detected in the interaction and $Q_e(E_i)$ is the amount of p.e. detected for the i-th Compton electron with energy transfer $E_i$. Equation (\ref{eq:Q_gamma}) has the same form for the random variables $Q_{\gamma}$ and $Q_e(i)$, and their average values $\overline{Q_{\gamma}}$ and $\overline{Q_e(E_i)}$ over a fixed pattern of Compton electrons~\footnote{In practical simulations, the expression (\ref{eq:Q_gamma}) is used to simulate a single event by summing up the individual contributions from the Compton electrons. In full simulations, the amount of light for each Compton electron is randomly generated for an energy release $E_e(i)$. On the other hand, if we are interested in the average light output of this particular pattern of Compton electrons, we can obtain this average signal by summing the average values of the individual Compton electrons, $\overline{Q_e(E_i)}$.}.
The variance of the signal $Q_{\gamma}$ is a sum of the variances for individual Compton electrons. For any pattern of Compton electron production, we can write:

\begin{multline}
Var(Q_{\gamma})=\sum_i Var(Q_\text{e}(i))=
\\
=\sum_i \big{[}(1+v_1)Q_\text{e}(i)+v(p)Q_\text{e}(i)^2+
\\
+(1+v(p))Q_1 Q_e(i)v_1^\text{int}\big{]}=
\\
=(1+v_1)Q_{\gamma}+v(p)Q_{\gamma}^2+
\\
+(1+v(p))Q_1Q_{\gamma}v_1^\text{int}.  
\label{eq:VarQ_gamma}
\end{multline}

When calculating  $\sum_i Q_e(i)^2$, we used the fact of statistical independence of the individual contributions. In other words, the contribution of transfer non-uniformity can be applied to the total light collected if the detector is uniform, i.e. if all the light produced by Compton electrons exhibits the same transfer properties. 

The $Q_{\gamma}$ values have their own distribution over the possible Compton electrons subdivision pattern, with an average gamma signal, $\overline{Q_{\gamma}}$.  To obtain the variance of the gamma response, we should perform calculations similar to those for the variance of the signal in the case of varying transfer efficiencies. We will obtain:

\begin{multline}
Var(Q_{\gamma})=
\\
=(1+v_1)\overline{Q_{\gamma}}+v(p)(1+v(Q_{\gamma}))\overline{Q_{\gamma}}^2+
\\
+v(Q_{\gamma})\overline{Q_{\gamma}}^2+
(1+v(p))Q_1\overline{Q_{\gamma}}v_1^\text{int}\simeq
\\
\simeq
(1+v_1)\overline{Q}_{\gamma}+(v(p)+v(Q_{\gamma}))\overline{Q_{\gamma}}^2+
\\
+(1+v(p))Q_1\overline{Q_{\gamma}}v_1^\text{int}   
\label{eq:VarQ_gamma}
\end{multline}

where $v(Q_{\gamma})=\frac{Var(\overline{Q_{\gamma}})}{\overline {Q_{\gamma}}^2}$ the variation of the mean $\overline{Q_\gamma}$ is calculated over all possible decompositions into Compton electrons, in the absence of other sources of smearing.

Both $v(p)$ and $v(Q_{\gamma})$ are small, so their product can be neglected. As can be seen, equation (\ref{eq:VarQ_gamma}) has the same form as that for electrons with the same intrinsic resolution parameter and with the quadratic term acquiring an extra term $v(Q_{\gamma})$. It should be noted that, while the parameter $v(p)$ of the "quadratic" contribution is independent of energy (the total amount of emitted light), the term $v(Q_{\gamma})$ depends on energy. It can be estimated using MC method if the light production curve is known: every time the Compton electron is generated, the corresponding average amount of light in accordance with calibration curve is added to the sum without simulating any other smearing. After many trials, an average value for the amount of light will be obtained, characterised by the coefficient $k_{\gamma}=\frac{Q_{\gamma}}{Q_e}$ and its relative variance $v(Q_{\gamma})=\big{(}\frac{\sigma_{Q_{\gamma}}}{Q_{\gamma}}\big{)}^2$. Here, $Q_{e}$ is the light amount for an electron of the same energy.

From the point of view of the analysis, the width of the scintillation line for gamma's and electrons with the same amount of detected light $Q$ will differ by a factor of $v(Q_\gamma) Q^2$. This factor is non-zero for any scintillator with non-linear energy response. This additional variance will apparently contribute to the internal resolution, together with the considered IR contribution for electrons.

\section{Absolute PMT calibration and other sources of systematics in IR measurements}

Accurate calibration of the PMT signal in p.e. units is crucial for correctly determining the statistical contribution to the energy resolution. Methods for the absolute calibration of PMTs operating in single photoelectron (SPE) mode have been discussed in several dedicated papers, see e.g.~\cite{PMTcalibration,Saldanha}. 

The absolute calibration of a PMT is primarily complicated by the physics of secondary electron production. In an ideal scenario, the electron amplification process would fully convert the initial photoelectron's energy into a cloud of secondary electrons—a process we refer to as normal photoelectron multiplication. In reality, however, some electrons transfer only a fraction of their energy to secondary emission. In extreme cases, an electron may elastically scatter off the first dynode surface without generating any secondary electrons. Such events produce an anode signal systematically weaker than that of normal multiplication. As a result, the SER spectrum develops a low-amplitude tail, its mean response shifts toward lower amplitudes (left of the main peak corresponding to normal p.e. multiplication), and the overall signal variance increases. More details on the subject can be found in~\cite{WrightBook}. A phenomenological model of the SER was proposed in~\cite{PMTcalibration}, consisting of a sum of Gaussian main peak truncated at zero and an exponential contribution representing underamplified signals. While this model provides valuable insight into the SER structure, it is not universally applicable.

I identified a systematic issue in PMT calibration, affecting nearly all measurements reporting the relative light yield (photoelectrons per MeV). Specifically, in \cite{Swiderski:intrinsic:2012},~\cite{Formozov:PhD}, the peak of the SER distribution is associated with one photoelectron, rather than the true mean value of the SER~\footnote{I prefer to use the term "main peak" to describe the feature. Additionally, I use "(main) peak position" instead of "most probable value" because, for spectra with a strong underamplified branch with exponential behaviour, the most probable value may approach zero.}. Furthermore, in some of these studies, the relative variance  $v_1$ of the SER is either calculated from the peak of the SER distribution, ignoring deviations from a Gaussian~\cite{Formozov:PhD}, or arbitrarily assumed to be $v_1=0.1$ without presenting any measurements~\cite{Swiderski:intrinsic:2012}. Additionally, the same research group in~\cite{Moszynski-Errors} attributes discrepancies between single-electron and multi-electron PMT calibrations as a "new source of error" in scintillation measurements. However, the actual cause is the substituting of the true mean and full variance of the SER with values derived solely from the "normal multiplication" peak in low-gain measurements.

Taking the issue of calibration  into account, I attempted to estimate the associated systematics for each set of measurements. The relevant information concerning the setup of the measurements considered in this paper is summarised in Table~\ref{Table:AllMeasurements}. All measurements except from those of Borexino and~\cite{BC408ep} are performed using cylindrical samples (cylindrical quartz vials in the case of LS). 

\begin{table*}
\centering
\renewcommand\arraystretch{1.4}
\small
\begin{tabular}{|c|c|c|c|}
\hline 
Reference & Scintillator & Geometry & PMT type  \tabularnewline
\hline 
\hline 
Swidersky~\cite{Swiderski:intrinsic:2012} & EJ301 & 51 mm $ \diameter \times$ 51 mm & Photonis XP5212B   \tabularnewline
\hline 
Swidersky~\cite{Swiderski:intrinsic:2012} & BC408 & 40 mm $\diameter \times$ 50 mm  & Photonis XP5500B  \tabularnewline
\hline 
Formozov~\cite{Formozov:PhD} & LAB+1.5 g/l PPO & 50 mm $\diameter \times$ 200 mm & Hamamatsu R6231-100    \tabularnewline
\hline 
Vo~\cite{BC408ep} & BC408  & 6x6x1 cm plate & Hamamatsu R6231-100 (x 4)\tabularnewline
\hline 
Roemer~\cite{EJ200} & EJ200  & 50 mm $\diameter  \times$ 50 mm & Photonis XP5500  \tabularnewline
\hline 
Deng~\cite{Deng} & LAB+2.5 g/l PPO & 35 mm $\diameter \times$ 25 mm  &  Hamamatsu CR160 \tabularnewline
  &  + 3 mg/l bis-MSB &   &   \tabularnewline
\hline 
Borexino~\cite{BrxSpectroscopy,Pablo_PhD} & PC+1.5 g/l PPO & $\diameter$~3~m   &  ETL9351 (x 2000)\tabularnewline
\hline 
\end{tabular}
\caption{The characteristics of the setups considered in our paper.}
\label{Table:AllMeasurements}
\end{table*}

\section{Measurements with BC408 and EJ301 commercial scintillators}

In this section I analyse the data from the publication~\cite{Swiderski:intrinsic:2012}, and all the information below originates from this article unless clearly stated  otherwise. Our main interest lies in the results of the measurement of the intrinsic resolution of commercial EJ301 liquid scintillator and BC408 plastic scintillator. The setup of the experiment was intended for the intrinsic resolution measurement. It provides an excellent light yield, which is the most important quantity in this kind of measurements, as discussed above. 

Measurements were performed using the Wide Angle Compton Coincidence (WACC) technique in the energy range of 10 keV up to 4~MeV. The scintillator sample was coupled to a PMT, and a high purity germanium detector was used to study the coincidence energy spectra. Measurements were carried out on a 40~mm diameter 50~mm high BC408 plastic scintillator coupled to an XP5500B PMT, and a 2” diameter 2” high EJ301 liquid scintillator sample coupled to an XP5212B PMT.

To perform the correct analysis, we will also need the data
on the dependence of light yield on energy. The corresponding measurement accompanies the main results in~\cite{Swiderski:intrinsic:2012} in the form of light yield non-proportionality plots. Note that the "intrinsic resolution" calculated in the paper of interest corresponds to the term $v_\text{int}(Q)=\frac {Q_1}{Q} v_{1}^\text{int}$ in our notation, and not to the accurate definition of the intrinsic resolution $v(N_\text{ph})$ from~(\ref{Birks}). This is obtained by subtracting the statistical contribution from the experimental values and represents the contribution of the intrinsic resolution to the full energy resolution. In practice, this is a more convenient definition, in what follows, by IR we will mean its contribution to the total energy resolution.

The IR measurement results for EJ301 LS, presented in Fig. 7 of [1], report the full width at half maximum (FWHM) in the 10$\div$447~keV energy range. I converted the FWHM values to 1$\sigma$ resolution by dividing them by 2.355 and then derived the relative variance as $v_\text{int}=(\sigma_\text{int}/2.355)^2$. Similarly, experimental uncertainties were  recalculated using the same dataset. The data in Fig. 7 of~\cite{Swiderski:intrinsic:2012} represent the intrinsic resolution variance, with the statistical component already subtracted. To enable more flexible analysis, I reconstructed the original relative variance of the signal by using the light yield (LY) and $v_1$ values from~\cite{Swiderski:intrinsic:2012}, assuming that the statistical component had been subtracted from the data, in line with the assumptions quoted in the paper, i.e. using the formula:

\begin{equation}
v(E)=v_\text{int}(E)+\frac{1+v_1}{LY\cdot E\cdot f_\text{NL}(E)} \label{v(E)}.
\end{equation}

I subtracted the uncertainty associated with $LY$ from the data uncertainties.

To check the compatibility of the experimental data with the model~(\ref{Birks3}) I fitted the data to a function derived from (\ref{Birks2}):

\begin{multline}
v_\text{int}(E)=\frac{1+v_1}{LY\cdot E\cdot f_\text{NL}(E)}+
\\
+v(p)+(1+v(p))\frac {f_\text{NL}(1~\text{MeV})}{f_\text{NL}(E)\cdot E} v_{1}^\text{int} \label{FitFunc}.
\end{multline}

As the IR data is presented as a function of energy, I substituted the ratio $\frac{Q_1}{Q}$ by its equivalent $\frac {f_\text{NL}(1~\text{MeV})}{f_\text{NL}(E)}$. Here $f_\text{NL}$ is the nonlinearity function from Fig.6 of the paper~\cite{Swiderski:intrinsic:2012}. The values in Fig.6~\cite{Swiderski:intrinsic:2012} are normalised to unity at 447~keV. To use the data in the analysis, I fitted it with a smooth curve and extracted the value for $f_\text{NL}(1~\text{MeV})$. All other data can be used in their original form. The uncertainty of the $f_\text{NL}(E)$ function was neglected in the analysis. In the intrinsic resolution extraction procedure, the term $v(p)$ ($\delta_p$ in the authors' notation) was omitted. As stated in the article: "The  ... term is linked to the light collection at the photocathode and the transfer of photoelectrons to the first dynode. It is assumed that these effects do not introduce substantial spread to the measured pulse width in modern PMTs." While the term is indeed small, the data on the right-hand side of the energy range clearly show a constant "supporting" term. This is why we left the term free in the analysis.

A straightforward fit of the data to the function~(\ref{FitFunc}) yields a very small value of $\chi^2$, with a reduced value $\chi^2/n.d.f.\simeq0.3$. This indicates that the errors are overestimated. In fact, the original uncertainties in Fig.7~\cite{Swiderski:intrinsic:2012} take two values: 10\% at lower energies and 7.5\% at higher energies. I suggest that this is a generic type of uncertainty assigned to the instrument. To obtain more realistic probabilistic estimates of the results, I scaled the resulting uncertainties by a factor of 0.5, yielding reduced $\chi^2$ values close to 1 for this set of measurements.

\begin{figure}[ht!]
\centering  
\includegraphics[width=1.1\linewidth]{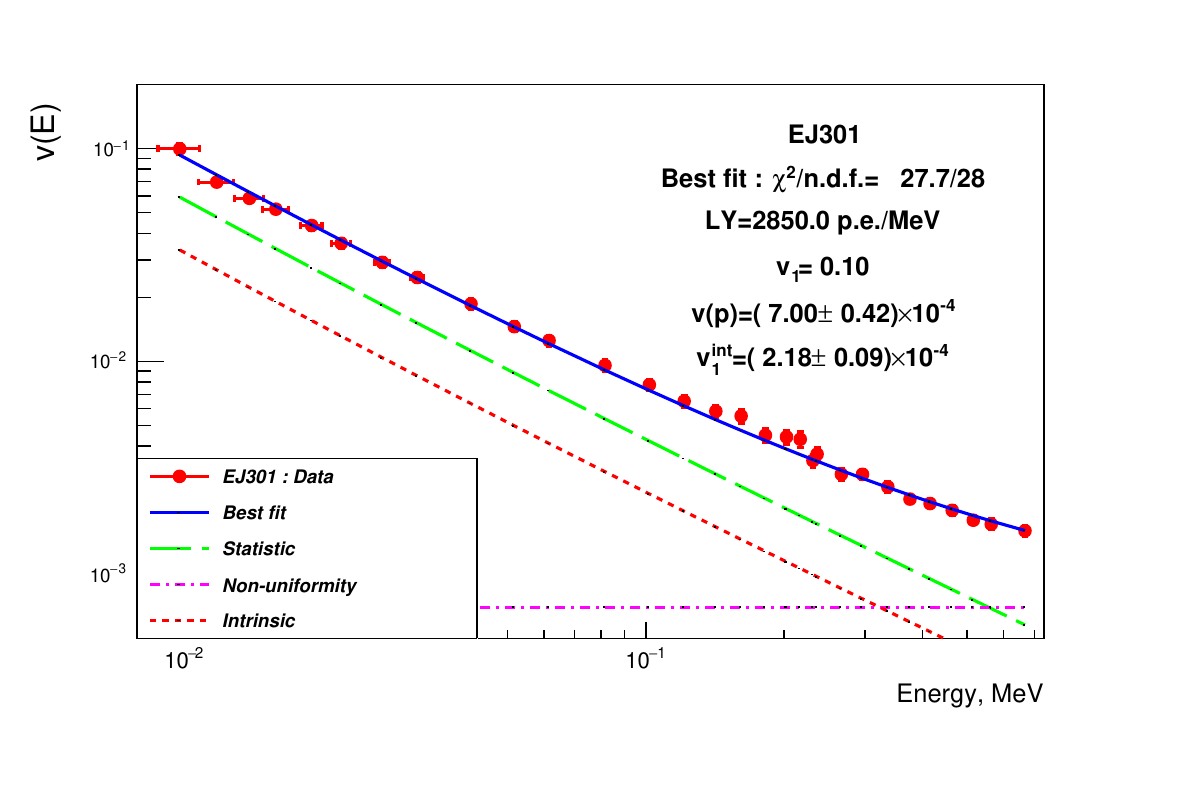}
\caption{The energy resolution of the EJ301 fitted using equation~(\ref{FitFunc}). The source of the data is Fig.7 of~\cite{Swiderski:intrinsic:2012}, the LY and $v_1$ parameters are fixed at the values provided in the paper~\cite{Swiderski:intrinsic:2012}. The figure shows the contributions of statistical fluctuations, IR and the (flat) contribution of light transfer non-uniformity.}
\label{Figure:EJ301:main_fit}
\end{figure}

Fig.~\ref{Figure:EJ301:main_fit} shows the data with the best-fit function superimposed.

The best-fit values are:

\begin{equation*}
\begin{array}{l}
v(p)=(7.00\pm0.42)\times 10^{-4},\\
v_1^\text{int}=(2.18\pm0.09)\times 10^{-4}.
\end{array}
\end{equation*}

with corresponding $\chi^2/n.d.f.=27.7/28$. The best-fit value for $v(p)$ corresponds to a reasonably low non-uniformity of the collected light of~2.6\%. When the value of $v(p)$ is fixed at "0", the $\chi^2$ value of the best fit is $\chi^2/n.d.f.=276.9/29$, which excludes the absence of the $v(p)$ term within the considered model.

Similar results are obtained for the BC408 plastic scintillator data extracted from Fig.4~\cite{Swiderski:intrinsic:2012}. The BC408 scintillator data includes a point at 4 MeV, which significantly extends the energy range of the EJ301 measurements. As in a previous analysis, fitting the data directly with equation~(\ref{FitFunc}) yields a very small value of $\chi^2$. To obtain more realistic probabilistic estimates of the results, we rescaled the uncertainties by the same factor 0.5, yielding reduced $\chi^2$ values close to 1 for this set of measurements.

\begin{figure}[ht!]
\centering  
\includegraphics[width=1.1\linewidth]{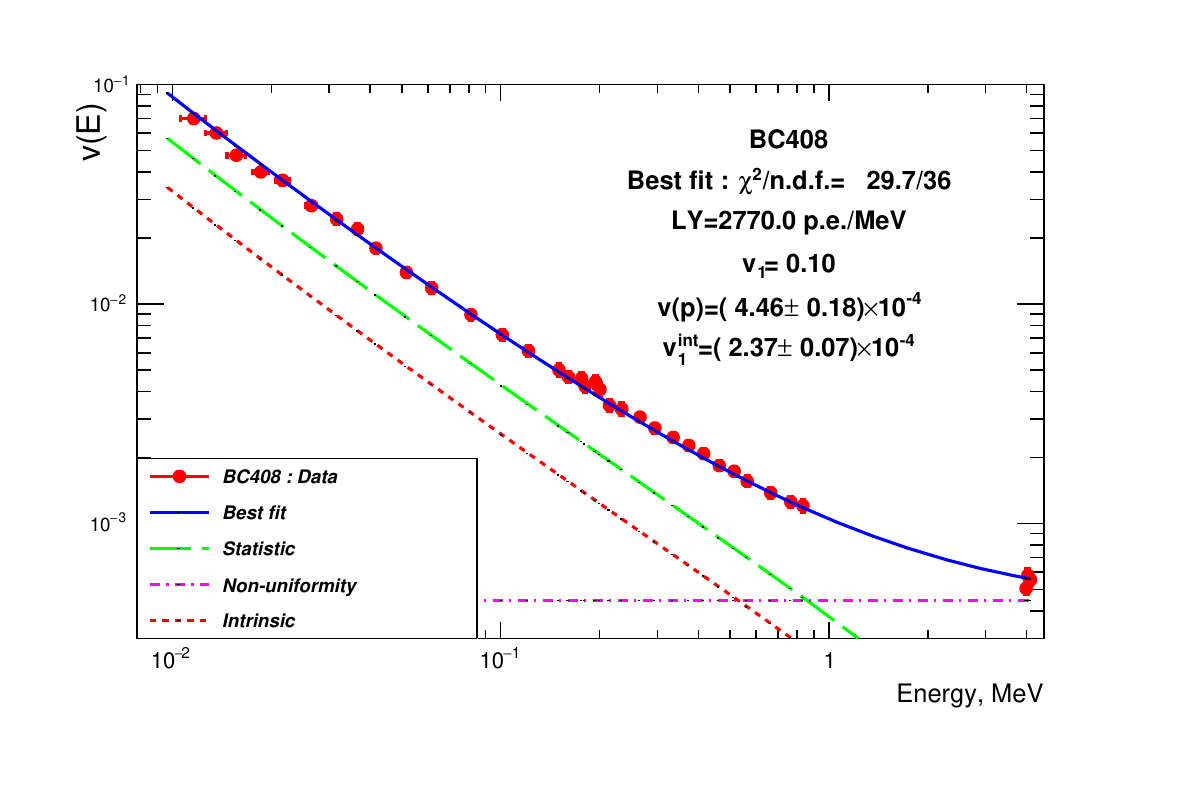}
\caption{The energy resolution of the BC408 fitted using~(\ref{FitFunc}). The source of the data is Fig.4 of~\cite{Swiderski:intrinsic:2012}), the LY and $v_1$ parameters are fixed at the values provided in the paper.}
\label{Figure:BC408:main_fit}
\end{figure}

The results are presented in Fig.~\ref{Figure:BC408:main_fit}, where we show the data with superimposed best-fit function. The best-fit values are:

\begin{equation*}
\begin{array}{l}
v(p)=(4.46\pm0.18)\times 10^{-4},\\
v_1^\text{int}=(2.37\pm0.07)\times 10^{-4}.
\end{array}
\end{equation*}

with the corresponding $\chi^2/n.d.f.=29.7/36$. The best-fit value for $v(p)$ is lower than in the previous fit; this could be explained by the difference in the setup. The value of $v(p)$ corresponds to 2.1\% non-uniformity of the collected light, which is just some fraction of a percent better than in the setup used for the EJ301 measurements. When the value $v(p)$ is fixed at 0, the $\chi^2$ of the best fit is $\chi^2/n.d.f.=622.3/37$, excluding the absence of $v(p)$ term, the same as in the EJ301 case.

We repeated the fit without the points at 4 MeV to check whether the result was stable across two energy intervals. The best-fit values in this case are as follows:

\begin{equation*}
\begin{array}{l}
v(p)=(5.16\pm0.35)\times 10^{-4},\\
v_1^\text{int}=(2.27\pm0.09)\times 10^{-4}.
\end{array}
\end{equation*}

with corresponding $\chi^2/n.d.f.=21.9/33$.

\section*{Discussion of the possible systematics of the measurement}

The main concern is the precision of description of the statistical term:
$$
v_{stat}\equiv c\cdot \frac{1+v_1}{Q}. \label{Stat_term}
$$

The possible bias in PMT calibration (i.e. deviation of the factor "c" from unity) is not discussed at all in the article~\cite{Swiderski:intrinsic:2012}. The uncertainty of the $Q$ measurement can be derived from the light yield estimate. The reported LY for the EJ301 scintillator is $2850\pm150$ p.e./MeV (5.4\%) and $2770\pm150$ (5.3\%) for the BC408 one. As discussed above, the single p.e. response was characterised using the position of the peak in the single p.e. in~\cite{Swiderski:intrinsic:2012}, thus introducing a bias in the measurement (i.e. $c\neq1$).
The peak position does not necessarily coincide with the true single p.e. average value. In the presence of a strong fraction of the underamplified~\footnote{In a book~\cite{WrightBook} these signals are called undersized, I prefer the term "underamplified"} signals, the difference could be significant. The same applies to the evaluation of the $v_1$ value from fitting a Gaussian to the peak. As will be shown below on an example of another measurement, the difference could be quite significant. The value of $v_1=0.1$ is mentioned for the case of the XP5212B PMT used in measurements with the EJ301 liquid scintillator. Apparently, the same value was used for the BC408 scintillator measurements with a different PMT, although this is not explicitly stated in the paper.

Taking these considerations into account in the absence of the real data on the SER, we checked the stability of the result with respect to the calibration value bias $c$ and relative resolution $v_{1}$. First, we fixed $v_{1}$ at 0.1, as reported in the paper, and set $c=1.1$ and evaluated $v_\text{int}(E)$ for this case applying the same fit to the data. Then, we set $v_1=0.2$ with $c=1.0$ and performed another fit. The results for the two scintillators are presented in Table~\ref{Table:Systematics}. The values of $\chi^2$ and $v(p)$ are not quoted in the table, as they are the same for all cases, because of the 100\% correlation between the parameters. As "c" and "LY" are correlated and are related to the same quantity, we made the test only for "c" values. Using the data from Table~\ref{Table:Systematics}, one can derive the following linear relations, connecting $v_1^\text{int}$ with the deviations from the central values of $v_1$ and $c$:

%\begin{equation}
%\begin{array}{l}
%\begin{strip}
\begin{multline}
v_1^\text{int}(\text{EJ301})=(2.18-3.6\cdot (c-1)
-
\\
-3.5\cdot (v_1-0.1))\times 10^{-4},\\
v_1^\text{int}(\text{BC408})=(2.37-4.0\cdot (c-1)
-
\\
-3.6\cdot(v_1-0.1))\times 10^{-4}.
%\end{array}
\label{Relations}
%\end{equation}
\end{multline}
%\end{strip}

Varying the light yield within its quoted uncertainty leads to a 9\% variation in the central values of $v_1^\text{int}$, yielding $0.18$ for EJ301 ($3.3 \times 0.054$) and $0.19$ for BC408 ($3.6 \times 0.053$). While the data presented in~\cite{Swiderski:intrinsic:2012} do not support additional conclusions, we note that systematic uncertainties associated with $LY$ and $v_1$ biases may be substantial. Typically, when calibrating the PMT single p.e. response using the peak rather than the full shape, the value of "c" is biased below 1 (see Fig.~\ref{Figure:XP5500}). The bias depends on the signal shape. Values of $v_1$ in the range $0.2\div0.3$ will systematically reduce the value of the parameter $v_1^\text{int}$, at the same time the corresponding calibration parameter will be below 1, which will partially compensate for the bias in the $v_1$ value. Assuming realistic values $v_1=0.25$ and $c=0.85$ we get almost the same values for $v_1^\text{int}$ for both scintillators. 

Taking this into account, we conclude that the uncertainty estimate of $v_1^\text{int}$ is dominated by systematics related to the definition of PMT parameters. A conservative estimate of the uncertainty is 10\%, as this aligns with the reported LY uncertainty.

\begin{table}
    \centering
\begin{tabular}{|c|c|c|c|}
\hline
 c & $v_1$ & EJ301: $v_1^\text{int}$ & BC408: $v_1^\text{int}$ \tabularnewline
   &  &  $\times{10^{-4}}$ & $\times{10^{-4}}$ \tabularnewline
\hline 
 1.0 & 0.1 & $2.18\pm0.09$ & $2.37\pm0.07$\tabularnewline
\hline 
 0.9 & 0.1 & $2.56\pm0.09$ & $2.77\pm0.07$\tabularnewline
\hline 
 1.0 & 0.2 & $1.83\pm0.09$ & $2.01\pm0.07$\tabularnewline
\hline 
\end{tabular}
\caption{Dependence of $v_1^\text{int}$ on SER relative variance $v_1$ and calibration bias $c$.}
\label{Table:Systematics}
\end{table}

An absence of intrinsic contribution ($v_1^\text{int}=0$) is only possible for unrealistically large values of $v_1$ and $c$. Using the relations (\ref{Relations}), the corresponding values are $v_1=0.73$ for EJ301, and $v_1=0.76$ for BC408 when $c=1$. Due to the calibration method (peak instead of the mean), we expect $c<1.0$, so there are no reasons for considering the values above $c=1$.  Therefore, we can conclude that zero intrinsic resolution contribution is excluded by the data.

We do not include the possible systematics due to the all discussed effects in the final result, and use only the uncertainty of the LY presented in the paper. The final results are compiled in Table~\ref{Table:AllIRData}.

We would like to emphasise the compatibility of the data with our intrinsic resolution model for any realistic set of $c$ and $v_1$ values, although the value of the parameter $v_1^\text{int}$ has additional systematic uncertainty because of the calibration bias and the lack of a reliable evaluation of the relative variance of the single-electron response, $v_1$.

\section{Measurements with the EJ200 plastic scintillator}

A set of measurements using an EJ200 plastic scintillator were reported in~\cite{EJ200}. The WACC technique was
applied to a $5\times5$ cm cylindric EJ200 plastic scintillator coupled to a Photonis XP5500 photomultiplier.

The non-linearity of the EJ200 response was measured in~\cite{EJ200}.  It fits well with the following function:
\begin{equation}
    f_\text{NL}(E)=-0.17\cdot E+0.39\sqrt{E}+0.80,
    \label{Eq:NL}
\end{equation}

where $E$ is energy measured in MeV. For our purposes, we renormalised the function to unity at 1 MeV by dividing $f_\text{NL}(E)$ by $f_\text{NL}(1)=1.02$. As the non-linearity was studied using a set of data with minimum energy of $E\simeq$~100~keV, doubts remain whether~(\ref{Eq:NL}) can be extrapolated to lower energies, given that at least four data points in the resolution plot are below 100~keV. Thus, we applied both the interpolation (\ref{Eq:NL}) and the results of the BC408 non-linearity measurements (an analogue of EJ200) from~\cite{Swiderski:intrinsic:2012} and found no differences.

For our analysis, we used the Compton electrons data from Fig.7 of~\cite{EJ200}, converting reported FWHM resolution to 1$\sigma$ resolution by dividing by 2.355. We set the relative errors at 10\%, this uncertainty is reported in the paper as the maximum relative error for the measured resolution values.

A straightforward fit using the expression (\ref{FitFunc}) apparently underestimates the values at low energies. This is due to the energy binning used in the analysis ($\Delta E$=10 keV) at the germanium detector side. To account for the width of the energy bin, we added the term $\frac{{\Delta E}^2}{12\cdot (f_\text{NL}\cdot E)^2}$ with a fixed $\Delta E$ to the expression  (\ref{FitFunc}), assuming a rectangular distribution of energy within the 10 keV energy bin.
The light yield of the setup is not discussed in the text of~\cite{EJ200}. The value of 2500 p.e./MeV is used in the caption of Fig.7~\cite{EJ200} for illustration purpose, but the "statistical term" in Fig.7~\cite{EJ200} is plotted without considering the PMT response width $v_1$. Because the setup used (a 50 mm diameter plastic scintillator with XP5500 PMT) is similar to the measurements with the BC408 considered above and taking into account that the manufacturers state the light output of 64\% anthracene for both scintillators, we can reasonably assume the same LY in both measurements, i.e. $2770\pm150$ p.e./MeV.

%The reported LY=2500 p.e./MeV contradicts the statistic contribution reported in Fig.7 of \cite{EJ200}(grey curve). Indeed, the FWHM resolution at 1 MeV with this p.e. counting would be $R_{FWHM}(\text{1 MeV})=\frac{2.355} {\sqrt{2500}}=4.71$~\% for an ideal PMT ($v_1=0$), while the value of $R_{FWHM}(\text{1 MeV})$ in Fig.7 corresponds to the lower value of 4.4~\% indicating the LY is underestimated. 

\begin{figure}[ht!]
\centering  
\includegraphics[width=1.1\linewidth]{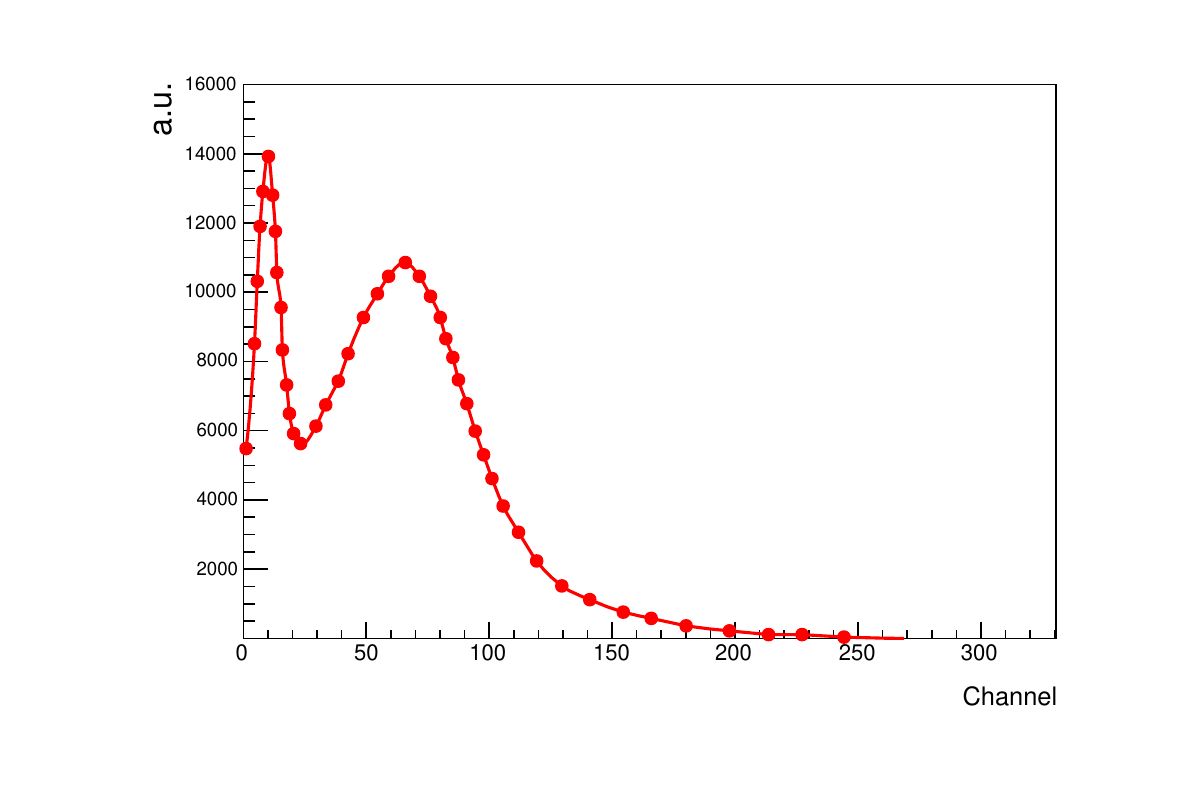}
\caption{The single electron response of PMT XP5500 from~\cite{XP5500} (Fig.~7, the one taken at +35$^{\circ}$ C).}
\label{Figure:XP5500}
\end{figure}

An example of the single-electron response (SER) of the Photonis XP5500 photomultiplier tube  is reproduced in Fig.~\ref{Figure:XP5500}. The SER spectrum exhibits a pronounced underamplified signal branch and a long tail of high-amplitude events, the latter are likely to be a contribution from multi-electrons response. No further experimental details or numerical estimates of SER parameters are provided in~\cite{XP5500}. To quantify the observed distribution, we calculated the mean charge $q_1=61.1$ and the relative variance $v_1=0.42$ for the full spectrum. The main peak was also fitted with a Gaussian, yielding $q_1^\text{peak}=63.8$ and $v_1^\text{peak}=0.20$. The minimal discrepancy between $q_1$ and $q_1^\text{peak}$ in the presence of a strong underamplified component suggests that the measured distribution does not represent a "pure" SER, but rather a response to a low-intensity light source with pedestal subtracted. To obtain a more realistic estimate of the parameters, we calculated $q_1=50.0$ from the distribution cut above 100, corresponding to $\simeq2\cdot q_1$. The corresponding $c$ value is then $c=0.78$ (50/63.8) and the light yield is $LY=3550$, noticeably high compared $2770\pm150$ p.e./MeV. Nevertheless, the new value is more close to the value of $LY=4050 \pm 200$ reported for a very similar setup with the same type of PMT~\cite{Nassalsky}. The relative variance of the truncated distribution is $v_1=0.31$ 

I also reconstructed a value for the relative variance, $v_1$, differently. First, I assumed that the spectrum consists of a peaked feature with a probability $p$, and that the remaining $(1-p)$ fraction do not contribute significantly to the signal, producing values with an average charge $q_\text{U}$ close to 0. The assumption is reasonable, as can be seen from Fig.~\ref{Figure:XP5500}. The true average value of the single electron spectrum is then:
$$
q_1=(1-p)\cdot q_\text{U}+p\cdot q_\text{peak}\simeq p\cdot q_\text{peak},
$$

where the $q_\text{peak}$ value corresponds to the average of the peak value. The variance of the single electron response is:

\begin{multline}
\sigma_1^2=<q^2>-<q>^2=
\\
=(1-p) \cdot q_\text{U}^2+p\cdot(q_\text{peak}^2+\sigma_\text{peak}^2) - q_1^2\simeq 
\\
\simeq p\cdot q_\text{peak}^2\cdot(1+v_\text{peak})-q_1^2,
\label{Eq:Sig1}
\end{multline}

where $\sigma_\text{peak}$ and $v_\text{peak}$ are the variance and relative variance, respectively, of the main peak in the single-electron response spectrum.

The relative variance of the single-electron response spectrum is then given by:

\begin{equation}
v_1\equiv \frac {\sigma_1^2}{q_1^2}\simeq \frac{1+v_\text{peak}}{p}-1.
\label{v1}
\end{equation}

Using the values of $p=c=0.78$ obtained from~(\ref{v1}), we get $v_1=\frac{1+0.04}{0.78}-1=0.33$, a value close to that obtained for the truncated distribution.

\begin{figure}[ht!]
\centering  
\includegraphics[width=1.1\linewidth]{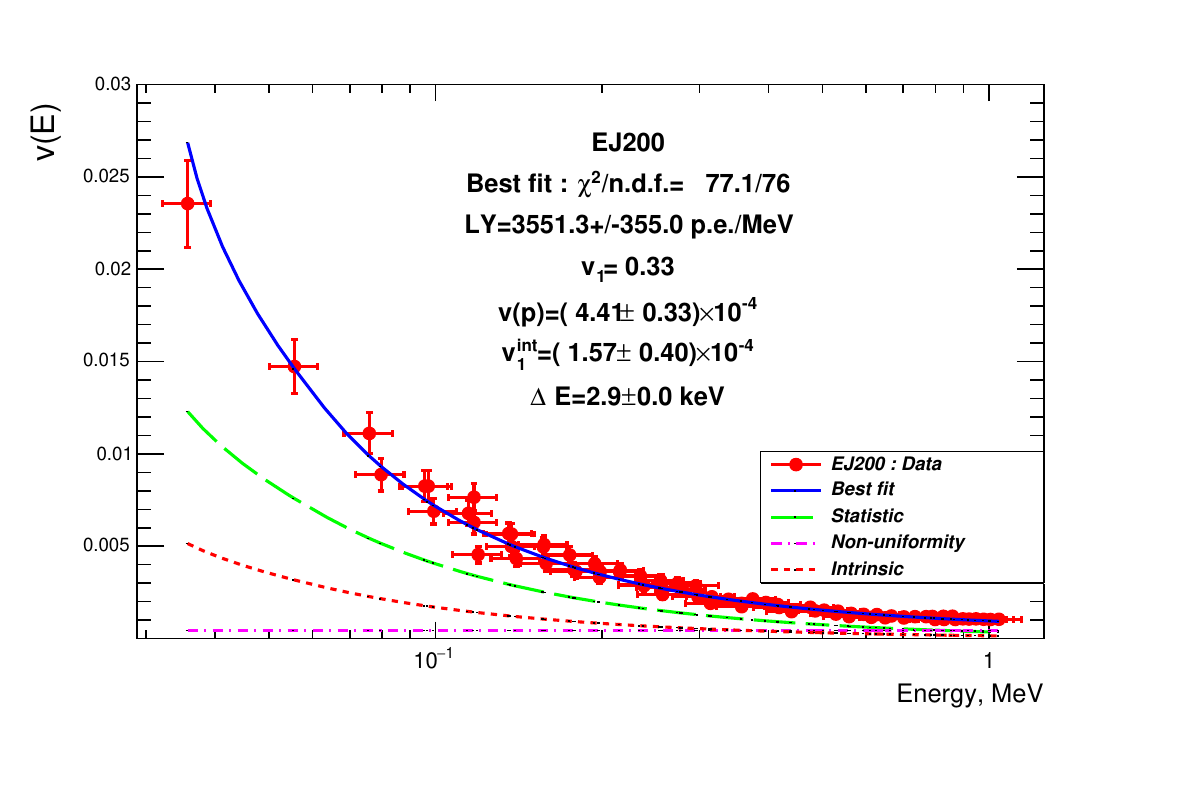}
\caption{The energy resolution of the EJ200 fitted with (\ref{FitFunc}). The source of data is Fig.7 of~\cite{EJ200}. The choice of the LY and $v_1$ parameters is explained in the text. The figure shows the contributions of statistical fluctuations, IR and (flat) contribution of light transfer non-uniformity.}
\label{Figure:EJ200}
\end{figure}

I set $v_1=0.33$ and increased the LY by 22\%, the approximate difference between the peak and the average value. I also accepted 10\% uncertainty for LY. The results of the fit are presented in Fig.~\ref{Figure:EJ200}.

\section{Advanced measurements with the BC-408 plastic scintillator}

In a paper~\cite{BC408ep}, the intrinsic resolution of the BC-408 plastic scintillator was estimated in a view of its potential application in searches for the neutrinoless double beta decay. As the energy range of interest for this type of physics is around~1 MeV, a careful study was performed within the relatively narrow range of 0.7 to 1.7 MeV. The response of the plastic scintillator to both the monoenergetic electrons and protons was investigated in this study.

The 6x6x1 cm BC-408 plastic scintillator plate under test was optically coupled from all sides to four square photomultiplier tubes (R6236-01, Hamamatsu). The monoenergetic electrons beam
was produced by an electron spectrometer capable to provide a range of electron energies from 0.7 MeV to 1.7 MeV (0.7, 1.0, 1.3, 1.5, and 1.7 MeV). The energy spread for 1.0~MeV
electron beam was measured to be approximately 1.8\% FWHM using high-resolution Si(Li) detector. As reported in~\cite{BC408ep}, this value includes contributions from the energy losses of the electron beam as it propagates through air and materials towards the Si detector, as well as the detector's resolution. 

The measurements in this study were conducted also with proton beam lines (microbeams) located at the WERC, Japan~(see references in~\cite{BC408ep}). Two specific energies were used, the energy spread for protons of 2.8~MeV and 3.4~MeV measured by the Si detector was confirmed to be less than 1\% (FWHM).

The measurement scheme involves recording the charges detected by each of the four PMTs in the set-up, denoted $Q_1,Q_2,Q_3$ and $Q_4$. Two combinations of these values are used in the analysis: the sum of all four charges, $Q^+=Q_1+Q_2+Q_3+Q_4$, and the differential signal $Q^-=(Q_1+Q_3)-(Q_2+Q_4)$. The $Q^-$ signal provides a measurement of the statistical part, which can be used to extract the internal resolution from the $Q^+$ values. In this scheme, the correlated contributions (internal resolution and the contribution of the electron beam energy spread) are mutually cancelled, providing a significant advantage over schemes with a single PMT. Another advantage of the four-PMT scheme is that it reduces the non-uniformity of the effect of single PMT responses by a factor of four. 

Indeed, the variance of the
total charge can be expressed as the sum over variances over all PMTs of the setup for an event:

\begin{equation}
\begin{split}
\sigma_{Q}^2=\sum_{i}^{N_{PMT}}(1+v_{{1}_i})\mu_{i}+v_i(p)\mu_{i}^2=
\\
=Q_0 \frac{1}{N_{PMT}}\sum_i^{N_{PMT}}(1+v_{1_{i}})+
\\
+Q_0^2\frac{1}{N_{PMT}}\sum_i^{N_{PMT}}v_i(p) =
\\
=Q_0 (1+\overline{v_{1}})+Q_0^2\frac{\overline{v(p)} }{N_{PMT}},
\end{split}
\label{Eq:Sig(Q)}
\end{equation}

where $N_{PMT}$ is the total number of (equal) PMTs in the setup. I ignored intrinsic resolution contribution for simplicity and used the following notations for average and relative variance of the involved quantities:

\begin{equation}
    \overline{v_1}\equiv \frac{1}{N_{PMT}}\sum_i^{N_{PMT}}v_{1_{i}},
\end{equation}

\begin{equation}
    \overline{v(p)}\equiv \frac{1}{N_{PMT}}\sum_i^{N_{PMT}}v_i(p).
\end{equation}

Thus, in the case the i-th PMTs exhibit non-uniformity $v(p)_i$, in the setup consisting of 4 PMTs the resulting contribution will be a quarter of the average, i.e. $v(p)=\frac{\overline{v(p)}}{4}$, where $\overline{v(p)}=\frac{\sum v(p)_i}{4}$.  

\begin{figure}
    \centering
    \includegraphics[width=1.1\linewidth]{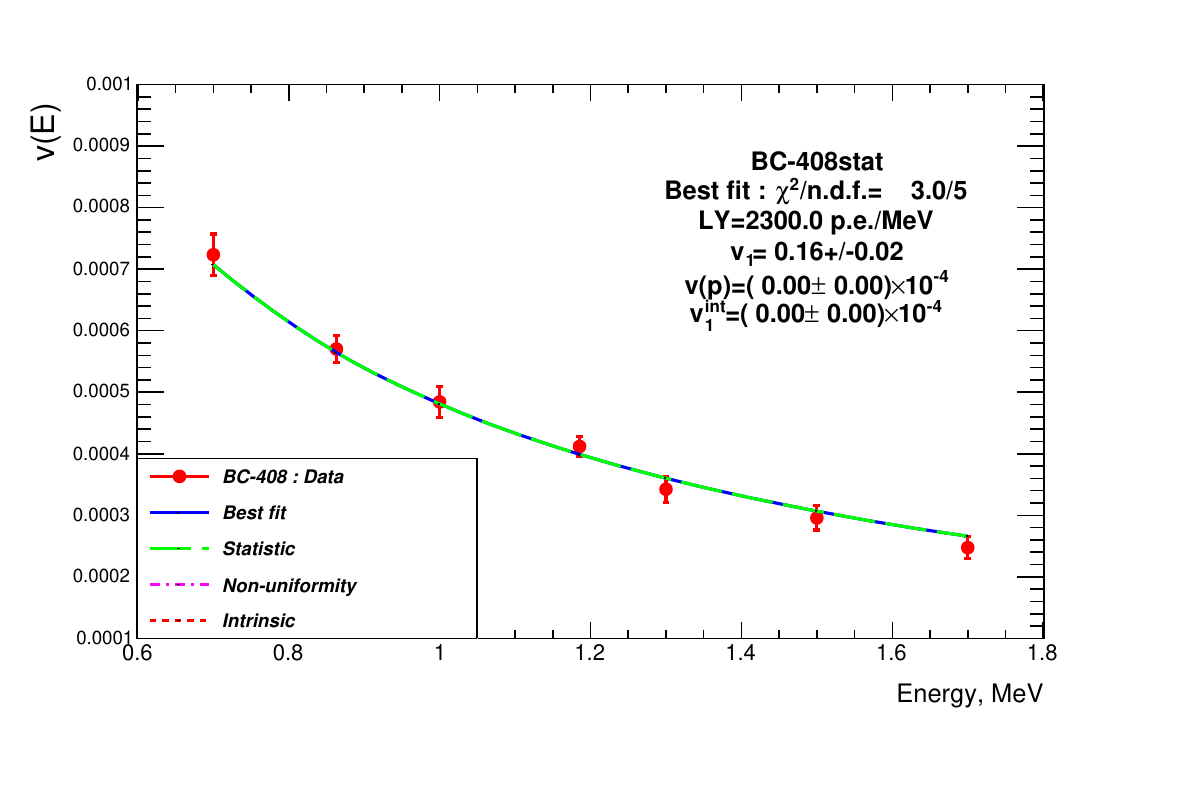}
    \caption{The fit of the statistical energy resolution data from~\cite{BC408ep} (see Fig.13 of~\cite{BC408ep}) with $\frac{1+v_1}{LY\cdot f_\text{NL}\cdot E}$ function.}
    \label{Figure:Fit:Vo:StatData}
\end{figure}

First, we fitted the pure statistical dataset ($Q^{-}$ in our notations). We fixed LY at the value of 2300 p.e./MeV and left $v_1$ free. It should be noted that, because of the correlation of these parameters, a precise description of the statistical part can be achieved using either of the two parameters involved. We chose LY value as reported in the paper. The non-linearity function $f_\text{NL}(E)$ is not provided in the paper, so we used the results from another set of measurements from~\cite{{Swiderski:intrinsic:2012}}
The best fit returns $v_1=0.162\pm0.012$, as shown in Fig.~\ref{Figure:Fit:Vo:StatData}.

At the next stage of the analysis, we observe that the residual energy resolution values for protons, once the statistical component has been subtracted, are at the level of $1\times10^{-4}$. As the design of the setup is very similar to previous cases, we can expect a similar contribution from the non-uniformity of the response on a single PMT. As shown above, the contribution from the signal summed from four PMTs will correspondingly be reduced by a factor of 4, providing a contribution at the level of $v(p)\simeq 10^{-4}$. Therefore, we can conclude that the contribution of intrinsic resolution for protons is $\simeq0$ and neglect it in the analysis.

\begin{figure}
    \centering
    \includegraphics[width=1.1\linewidth]{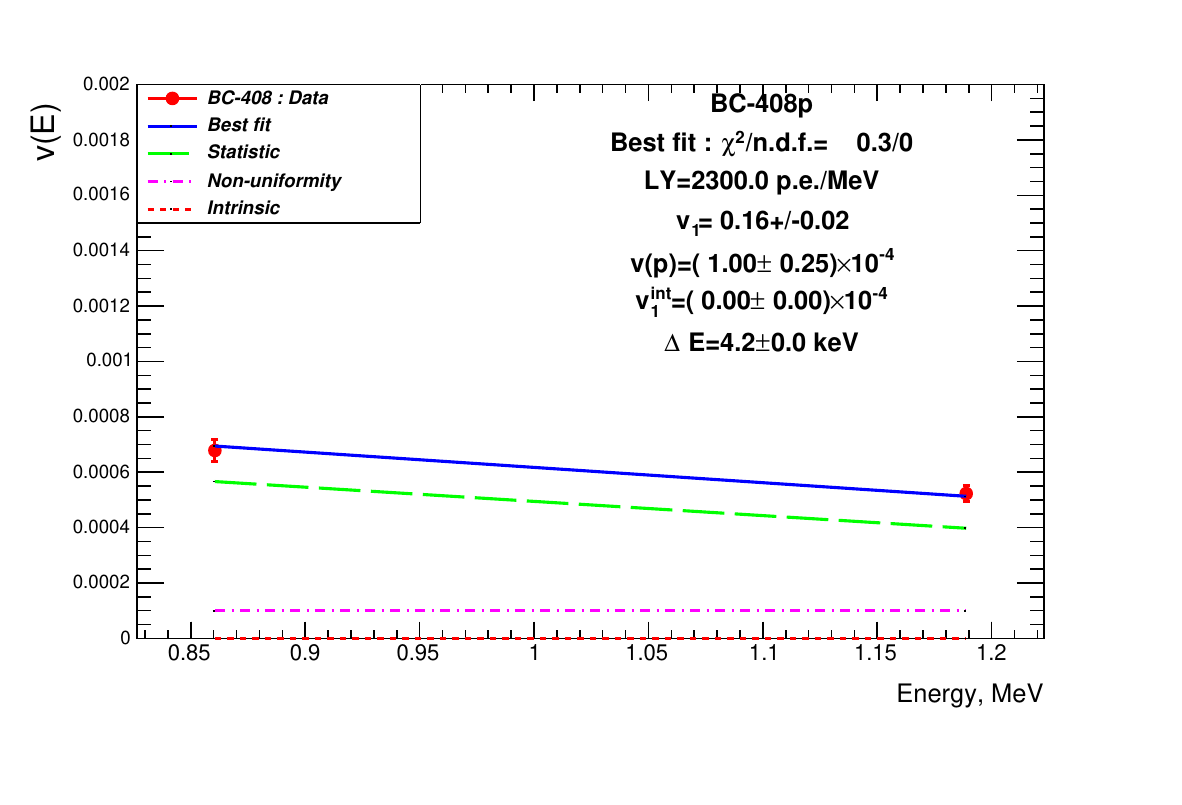}
    \caption{The fit of the resolution data (see Fig.14 of~\cite{BC408ep}) for proton beam  with constrained statistical term and free $v(p)$. The smearing of the proton beam is taken into account as a penalty term in the $\chi^2$.}
    \label{Figure:Fit:Vo:protonsData}
\end{figure}

The fit of the proton energy resolution data is presented in Fig.~\ref{Figure:Fit:Vo:protonsData}. The best-fit value for $v(p)=(1.00\pm 0.25)\times 10^{-4}$, which agrees with the expected 1/4 of the single PMT case.

\begin{figure}
    \centering
    \includegraphics[width=1.1\linewidth]{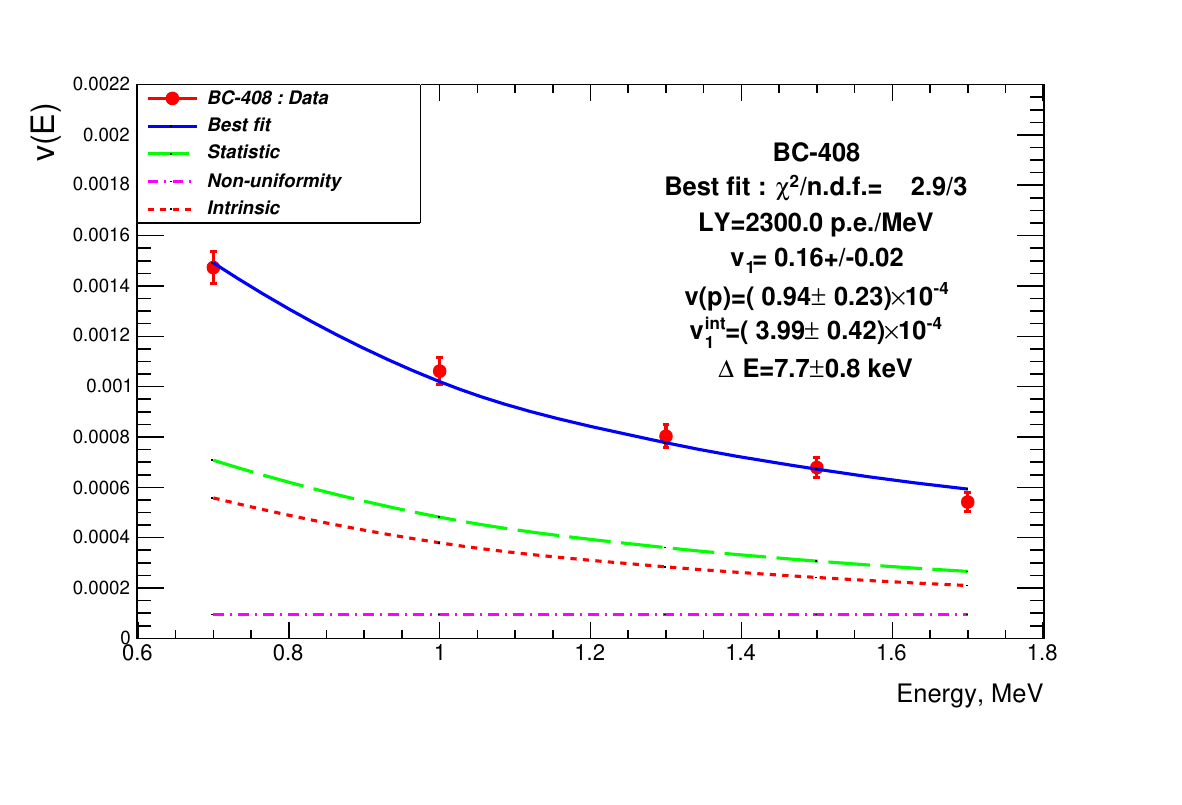}
    \caption{The fit of the data (see Fig.14 of~\cite{BC408ep}) for electrons from~\cite{BC408ep} with constrained statistical term and $v(p)$. The smearing of the electron beam is also taken into account as a penalty term in the fit.}
    \label{Figure:Fit:Vo:elData}
\end{figure}

The data on the energy resolution for electrons can now be processed using the statistical contribution parameterisation and the $v(p)$ value from the protons data analysis. We also took into account the reported beam smearing: 1.8\% FWHH at 1 MeV. The results are presented in Fig.~\ref{Figure:Fit:Vo:elData}. The uncertainty of the statistical term subtraction (described by the $v_1$ parameter in this case) propagates to 10\% uncertainty of the final result on $v_1^\text{int}=(3.99\pm0.42)\times 10^{-4}$. The value of $\Delta E=7.7$~keV corresponds to 1.8\% FWHH at 1 MeV, and we assume it is constant across the energy range of interest.

To test the assumption that the intrinsic resolution contribution for protons is negligible compared to the total resolution, I performed a more sophisticated analysis incorporating both proton and electron datasets.

In this analysis, the same values for light yield ($LY$), $v_1$, and $v(p)$ were assumed for both electrons and protons. The beam widths were fixed to the measured values provided in~\cite{BC408ep}. The $\chi^2$-profile for the proton intrinsic resolution ($v_{1_{p}}^\text{int}$) was obtained by performing a fit over a grid of $v_{1_{p}}^\text{int}$ values for protons, while leaving $v_{1_{e}}^\text{int}$ for electrons as a free parameter.

The light yield (LY) was fixed to the value obtained from the statistical data. The $v_1$ parameter was constrained to its best-fit value from the same dataset. The $v(p)$ parameter was constrained in the fit to $v(p) = 1.00 \pm 0.25$, a value derived from a fit to the proton data with both $v(p)$ and $v_{1_{p}}^\text{int}$ left free. This value corresponds to a range of $v(p) = 3.0 \div 5.0$ for a single PMT, which is consistent with values observed in previously considered setups.

No significant bias in the estimate of $v_{1_{e}}^\text{int}$ for electrons was observed. Furthermore, the shape of the $\chi^2$-profile excludes values of $v_{1_{p}}^\text{int}$ for protons above $0.4\times10^{-4}$ at 90\% C.L..

The measurement in~\cite{BC408ep} are the most reliable among the all measurements reviewed in this paper, as the absolute calibration of PMTs is not an issue. We also note the robust estimate of the measurement uncertainties compared to that of other considered measurements. 

\section{Measurements with the LAB+PPO liquid scintillator}

The results of another measurement using a monoenergetic $\gamma$-source with WACC were reported in~\cite{Formozov:2019}. The contribution of the intrinsic resolution was estimated to be $v_\text{int}(50\text{ keV})=0.02\pm0.005$. 

Using our model, we calculated the value of $v_1^\text{int}$: $v_\text{int}=\frac{Q_1}{Q}v_1^\text{int}=\frac{LY\cdot 1}{LY\cdot 0.05\cdot f_\text{NL}(E,kB)}v_1^\text{int}\simeq \frac{1}{0.05\cdot0.9} v_1^\text{int}$, we get $v_1^\text{int}=(9.0\pm2.3)\times 10^{-4}$, which corresponds to an intrinsic resolution of $(3.0\pm0.4)$\% at 1~\text{MeV}~\footnote{the values in our paper~\cite{MyIntrinsic} are different: $v_1^\text{int}=(11\pm3)\times^{-4}$ and $R_\text{int}(\text{1 MeV})=(3.3\pm0.4)$\%, this is the result of a miscalculation, division by 0.9 instead of multiplication.}. The value $f_\text{NL}(k_\text{B},E_\text{e}=0.05\text{ MeV})\simeq 0.9$ is obtained from Fig.3 of~\cite{Formozov:2019}.

Below, we reanalyse the data using the additional measurement details described in the PhD thesis~\cite{Formozov:PhD}. 

The statistical term $\frac{1+v_1}{Q}$ in the PhD was described using $v_1=0.05$. The amount of detected p.e.  was characterised by the relation $Q=LY\cdot f(k_\text{B},E)\cdot E$. The ionization quenching parameter $k_\text{B}=0.0196$ cm/MeV was measured independently. The measured light yield was $LY=450$ p.e./MeV, which is significantly lower compared to the previously discussed measurements. When attempting to fit the data in a manner similar to that used in the previous analysis, it became evident that the dataset with the subtracted statistical term still exhibited dependence on energy within the considered energy range: the dependence of data on energy in the considered energy range is very weak. Comparing the data for the intrinsic contribution with the data for EJ301/BC408, it is evident that the values are too high. Upon closer examination of the statistical term subtraction procedure, we found that the value of the $v_1$ used in analysis was too low. Indeed, the Hamamatsu type R6231 100 PMT used for the measurements doesn't demonstrate such excellent single-electron response resolution.

The fit of the spectrum obtained using a low-intensity light source is shown in Fig.2.29 of~\cite{Formozov:PhD}. The data in the plot for two-photoelectrons contribution can be used to evaluate it variance as $v_2\equiv\frac{p5}{p4}=(\frac{0.0758}{0.4246})^2=0.032$, here p4 and p5 are parameters of the fit corresponding to the mean and variance of the double p.e. peak. However, the relative variance of the single photoelectron spectrum is factor 2 larger, i.e.  $v_1=0.064$, this already contradicts the $v_1=0.05$ value used in the analysis. 

Other available measurements of the Hamamatsu type R6231-100 PMT show the presence of a strong branch of underamplified signals in its single p.e. spectrum (see Fig.7 from~\cite{deHaas:PMT:2011}). The values of $v_1$, obtained for the two PMT samples in the aforementioned paper, are  $v_1=0.27$ and $v_1=0.28$. This feature is invisible in Fig.2.29 of~\cite{Formozov:PhD} because of the high threshold set in the measurements.
We note that, although the article~\cite{deHaas:PMT:2011} does not present an estimate of measurement errors, accurately determining the parameter $v_1$ is a challenging task for spectra with a significant contribution from underamplified signals due to the parameter's high sensitivity to spectral shape features.

The presence of a strong underamplified signal branch in the single-electron spectrum will also influence the average SER value: the peak value will shift above 1 p.e., while in the analysis in~\cite{Formozov:PhD} the position of the single p.e. is chosen practically at the peak value ($q_\text{peak}=0.217$ while 1 p.e. corresponds to 0.2$\div$0.21 in $V\cdot ns$ scale~\cite{Formozov:PhD}). The values of $c$ and $v_1$ can be reconstructed using (\ref{Eq:Sig1}) and (\ref{v1}).

Using the values $v_\text{peak}=0.05$ and $v_1=0.27(0.28)$, we can estimate the value of $p$ from equation (\ref{v1}): $p=0.83(0.82)$. Because of the definition of the calibration correction, the corresponding calibration bias coincides with p ($c=p$). This may seem surprising, but the new set of values practically does not influence the result. Indeed, instead of the factor $c\cdot(1+v_1)=1\cdot (1+0.05)$ we used the new factor $0.83\cdot (1+0.27)=1.054$ that practically coincides with the original factor. This occurs primarily because of the assumed SER structure, which enables neglecting the contribution from underamplified signals. The value of the parameter $v_1^\text{int}$ will not change because the ratio $\frac{Q_1}{Q}$ does not depend on the calibration.
A more careful analysis could reveal larger deviations. However, as previously mentioned, in the absence of an exact measurement of the PMT SER there is no opportunity to make final conclusions. We should also bear in mind the potential for additional systematic uncertainties in extracting $v_1^\text{int}$ value. 

In the fit, we used $v_1=0.27$. The corresponding calibration bias value is $c=0.83$ for the extreme case of zero contribution of the underamplified branch to the average value. To take the uncertainty of these values into account, we used $c=92$ (the average of 0.83 and 1.0), and assigned an absolute uncertainty on $c=0.92\cdot\frac{0.17}{\sqrt{12}}=0.05$, assuming the rectangular p.d.f. of the calibration values between the extremes $c=0.83$ and $c=1.0$.

Due to the small size of the PMT one would generally expect low non-uniformity of the light collection, $v(p)$. Similar to the measurements considered before, the value of $v(p)$, was not evaluated in~\cite{Formozov:PhD}. Because of the energy interval used, the precise value of $v(p)$ is unimportant, as the data are insensitive to its value within a fairly large range.
We set $v(p)=5\cdot10^{-4}$, as with the other measurements considered. This value falls within the margin of error of the data points.

The fit to the data was performed using the non-subtracted data from Fig.2.40 of ~\cite{Formozov:PhD}. The errors for the data points in $v(E)$ plot are $10\times10^{-4}$. However, we found that the uncertainties were overestimated and that the fit to the data returned too low values for the reduced chi-squared. To achieve a more realistic description of the data, we scaled the errors by a factor of 0.4.

\begin{figure}
    \centering
    \includegraphics[width=1.1\linewidth]{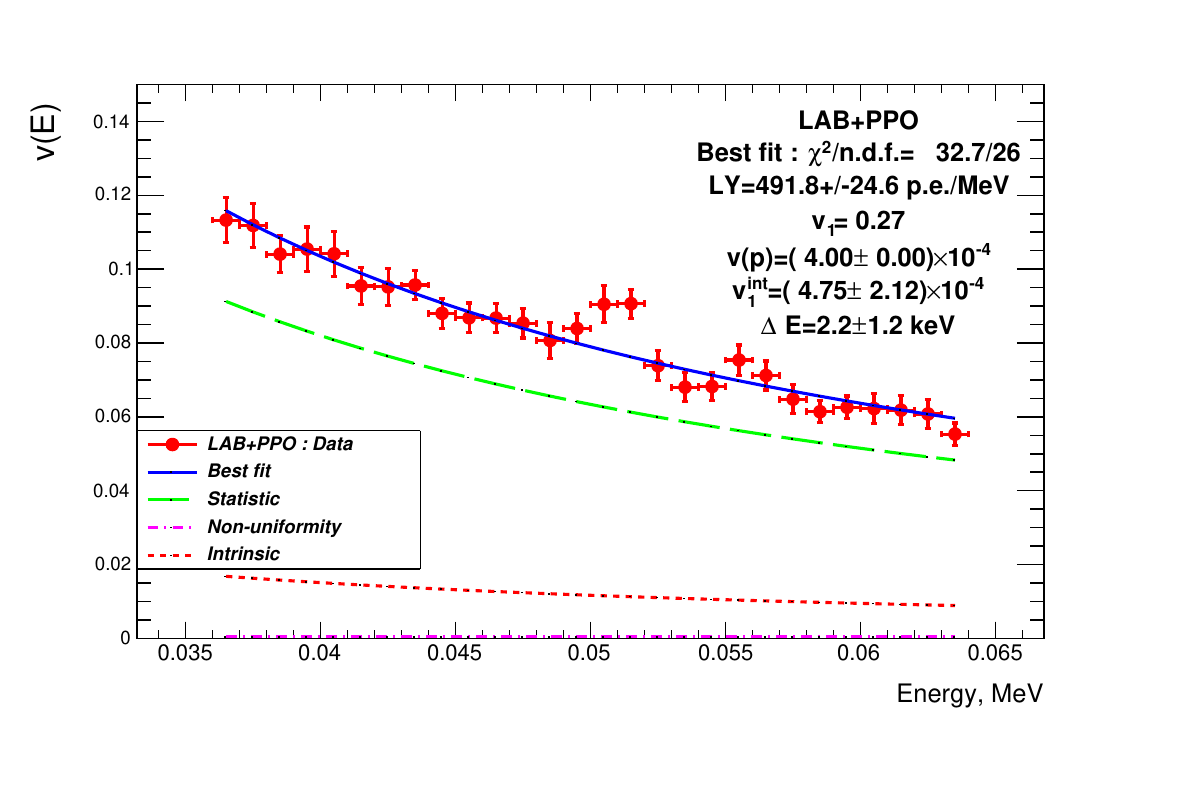}
    \caption{The fit of the data from~\cite{Formozov:PhD} with (\ref{Eq:Fit:Formozov})}.
    \label{Figure:Fit:Formozov:Data}
\end{figure}

The fit was performed using the function (\ref{FitFunc}) with additional term responsible for final energy width of the electrons sample as discussed in section~\ref{StatDesc}:

\begin{multline}
    v_\text{int}=v(p)+c\cdot\frac{1+v_1}{450\cdot E\cdot f(0.0196,E)}+
    \\
    +\frac{v_1^\text{int}}{f_\text{NL}(E)\cdot E}+\frac{\Delta E^2}{12\cdot (f_\text{NL}\cdot E)^2},
    \label{Eq:Fit:Formozov}
\end{multline}

where $f_\text{NL}(E)=f_\text{NL}(k_B=0.0196,E)$, and $\Delta E$ is the energy smearing in the Ge detector. I assume rectangular distribution of the electron energies withing the energy bin $\Delta E$. The fit returns:

\begin{equation}
    v_1^\text{int}=(4.8\pm2.1)\times10^{-4},
\label{Eq:Formozov:Results:450}
\end{equation}

with $\chi^2=32.7/26$ and $\Delta E=(2.2\pm1.2)$ keV, see Fig.~\ref{Figure:Fit:Formozov:Data}. The value of $v(p)$ does not depend on the choice of the set of the parameters $v_1$ and $c$, so we can conclude that at considered energies $v(p)$ is negligible in this experimental setup, namely it is $v(p)<6.0\times10^{-4}$ at 90\% C.L. as results from the $\chi^2$ profile study. The best-fit value of $\Delta E$ is compatible with the stated width of the energy bin on the Ge detector side of 1~keV, the effect of the Ge detector resolution probably contributes to this value as well.

The obtained value  $v_1^\text{int}=(4.8\pm2.1)\times10^{-4}$ corresponds to $R_\text{int}(\text{1~MeV})=(2.2\pm0.5)$~\%.

\section{The measurement with LAB-based LS using monoenergetic electrons at 1 MeV}

In this section, I examine the measurements reported in~\cite{Deng}. In our previous paper~\cite{MyIntrinsic}, I accepted the reported result without critical assessment. However, as the estimated uncertainties seem too low (see Table~3 of the paper~\cite{Deng}), I investigated the reliability of the presented errors. At first glance, it can be seen that the maximum reported uncertainty in the table is $0.11$\%, while the resulting uncertainty is $0.06$\%. This appears to be the result of a miscalculation, as the absolute uncertainties should be summed quadratically.

The systematic error, $\delta_\text{system}$, is estimated using a LaBr$_3$ scintillator. First, the possible contribution of the internal resolution of the LaBr$_3$ is not subtracted. Secondly, the result contains a non-uniformity contribution ($\delta_\text{uniformity}$, or $v(p)=(\delta_\text{uniformity})^2$ in our notations). This means that the non-uniformity contribution is subtracted twice in the result: once  directly, and again together with the $\delta_\text{system}$.

The PMT used for the measurements was calibrated using the peak value of the charge spectrum, as the reported resolution of a single- photoelectron (about 20\%) is unrealistically good. Although there are no plots of SER in the paper, but a very similar measurement was presented by a subgroup in another paper~\cite{Feng}, which provides the opportunity to reanalyse the spectrum. The SER spectrum is presented in Fig.9 of the paper~\cite{Feng}. One can observe a cut pedestal and a discrepancy between the mean values of the two fitted Gaussians: $\text{mean}_1=123.4$ and $\text{mean}_2=264$. If the two Gaussians represent the single electron response and two-electrons responses, then we should have $\text{mean}_2=2\cdot \text{mean}_1$. It appears that the pedestal position is not being taken into account. A rough estimate can be obtained from $\text{pedestal}=\text{mean}_1-(\text{mean}_2-\text{mean}_1)=-17.2$. 

The spectrum in Fig.9~\cite{Feng} represents full charge response spectrum shifted by -17.2 and cut at zero. It can be shown that the average ($\overline{q}$) and variance ($v(q)$) of the charge spectrum for the distribution with cut pedestal are as follows:

$$
\overline{q}=\frac{\mu}{1-e^{-\mu}}q_1,
$$
$$
v(q)=\frac{1-e^{-\mu}}{\mu}(1+v_1)-e^{-\mu}.
$$

In the case of a Poisson distribution of probabilities of detecting a fixed amount of photoelectrons with an average $\mu$,
the probability ratio of a two-electron response to a single electron response is given by $\frac{P(2)}{P(1)}=\frac{\mu}{2}$. The ratio of the corresponding areas, calculated from  Fig.7, gives us $\frac{P(2)}{P(1)}=0.348$, and hence we can estimate $\mu \simeq 0.695 \pm 0.044$. The uncertainty is estimated from the (unfitted) fraction of signals in the SER valley., 

Using the mean and r.m.s. data from the figure, we derive the following: $q_1=\frac{(162.5+17.2)\cdot(1-e^{-\mu})}{\mu}=129.3$ ($\overline{q}=162.5$ from Fig.7) and $v(q)=(\frac{96.86}{162.5+17.2})^2=0.290$  ($r.m.s.=96.86$ from Fig.7). The calibration parameter $c$ can be calculated using the values from Fig.7: $c=\frac{\text{mean}_1}{q_1}=0.92$. Similarly, I estimate $v_1 =0.10\pm0.01$ using $v(q)=0.290$. The former should be compared to the reported value of $v_1=0.04$, which corresponds to a very low 20\% SER resolution.
Using these values and taking into account the reported LY=3266 p.e., we obtain for statistical contribution of $\delta_\text{st}=(1.68\pm0.08)$\% (instead of 1.75\%), and for $\delta_\text{s.p.e.}=(0.55\pm0.11)$\% (instead of 0.35\%). The uncertainties of these values are dominated by the uncertainty in extracting $\mu$ value. Let me point out that the possible contribution of the internal resolution of the LaBr$_3$ detector is neglected in the estimate of systematic uncertainties in~\cite{Deng}; this point requires clarification. In the absence of dedicated measurements, we will take this fact as granted.

The recalculated value of the internal contribution at 0.976 MeV is $R_\text{int}(0.976~\text{MeV})=(1.85\pm0.18)$\%. Note the larger uncertainty and the fact that the central value has hardly changed compared to the value reported in~\cite{Deng}. Excluding the double count of the non-uniformity contribution from the final result increases the value of the internal contribution to $R_\text{int}(0.976)=(2.06\pm0.14)$\%. The latter value will be used to derive a corrected value of $v_1^\text{int}$ for this measurement.

\section{Discussion}

All data are consistent with our simple model within the measurement precision. Note that in a number of cases, the reported uncertainties were rescaled as they appeared to be overestimated. Furthermore, the data considered do not allow us to draw any conclusions regarding the presence of unaccounted additional fluctuations.

The values of the parameter $v_1^\text{int}$ obtained in present analysis, along with results derived in~\cite{MyIntrinsic} from Borexino publications, are compiled in Table~\ref{Table:AllIRData}. 

The data indicate an additional broadening of the energy resolution due to intrinsic resolution, contributing approximately $\simeq(1.5\div2)$\% at 1 MeV for all considered scintillators. For the BC408 (EJ200) plastic scintillator, different experiments report IR contributions in the range $\simeq(1.25\div2.0)$\% at 1 MeV. The higher values correspond to measurements around 1~MeV, while lower-energy data yield smaller values.  This discrepancy may arise from unaccounted systematics in low-energy measurements, particularly since the results in~\cite{BC408ep} are free of such systematics. Alternatively, it could suggest different underlying processes in the scintillator response below and above 1~MeV, leading to deviations from the phenomenological model above 1~MeV. 

For LAB-based liquid scintillators, the two measurements are consistent within measurement errors. The results from~\cite{Formozov:2019} agree with the more precise measurement by Deng et al.~\cite{Deng}, though with larger uncertainties.

The two Borexino measurements for PC-based LS show good agreement. The value reported in the more recent Borexino publication~\cite{BrxSpectroscopy} is likely more precise, as the parameterisation incorporates the energy dependence of the IR contribution. The earlier analysis yields an effective value; converting this into a universal parameter would require an accurate determination of the effective energy corresponding to the tail of the $^{14}$C spectrum above the analysis threshold. We further note that this result is unaffected by calibration bias and $v_1$ uncertainties, as it was obtained using the PMT occupancy (the number of triggered PMTs per event) rather than the total charge. See~\cite{MyIntrinsic} for a detailed description of the measurement methodology.

Notably, LAB-based scintillators exhibit a slightly higher IR contribution compared to PS-based ones.

Additionally, measurements with electrons at $\sim 1$~MeV~\cite{Deng,BC408ep} systematically yield higher IR values than those at lower energies, potentially deviating from the phenomenological model at higher energies. However, systematic measurement errors cannot be ruled out as the cause of this discrepancy, which stands at approximately 
$3\sigma$. More precise measurements across a broader energy range are needed for definitive conclusions.

Finally, an experiment with a monoenergetic proton source shows no intrinsic smearing effect, a finding that may be significant for the calibration of organic scintillator detectors.

\begin{table*}
\centering
\renewcommand\arraystretch{1.4}
\small
\begin{tabular}{|c|c|c|c|c|}
\hline 
 & Scintillator & Energy & $v_1^\text{int}$ & $R_\text{int}(1\text{  MeV})$ \tabularnewline
& & range& $\times 10^{-4}$ & \% \tabularnewline
\hline 
\hline 
Swidersky~\cite{Swiderski:intrinsic:2012} & EJ301~(C$_6$H$_4$(CH$_3$)$_2$)  & $10\div667$~keV  & $2.18\pm0.26$ & $1.48\pm0.08$\tabularnewline
\hline 
Swidersky~\cite{Swiderski:intrinsic:2012} & BC408~(C$_6$H$_4$(CH$_3$)$_2$)  & $10\div 667$~keV  & $2.27\pm0.27$ & $1.51\pm0.09$\tabularnewline
\hline 
Swidersky~\cite{Swiderski:intrinsic:2012} & BC408~(C$_6$H$_4$(CH$_3$)$_2$)  & $10\div 4000$~keV  & $2.37\pm0.27$ & $1.54\pm0.09$\tabularnewline
\hline 
Roemer~\cite{EJ200} & EJ200 (analog BC408) & $36\div 1040$~keV  & $1.57\pm0.40$ & $1.25\pm0.16$\tabularnewline
\hline 
Vo~\cite{BC408ep} & BC408~(C$_6$H$_4$(CH$_3$)$_2$)  & $0.7\div 1.7$~MeV  & $3.99\pm0.42$ & $2.0\pm0.1$\tabularnewline
\hline 
Formozov~\cite{Formozov:PhD} & LAB+1.5 g/l PPO & $35\div 65$ keV & $4.8\pm2.1$ &$2.2\pm0.5$  \tabularnewline
\hline 
Deng~\cite{Deng} & LAB+2.5 g/l~PPO   & $(1.83\pm0.06)$\%  & $4.26\pm0.59$ & $2.06\pm0.14$\tabularnewline
  &  + 3 mg/l bis-MSB &  @ 0.976 MeV &   & \tabularnewline
\hline 
Borexino~\cite{BrxSpectroscopy} & PS + 1.5 g/l PPO & $v(N)=(11.5\pm 1)$& $2.2\pm0.2$ & $1.5\pm0.1$\tabularnewline
  &   &  @ 156 keV &   & \tabularnewline
\hline 
Borexino~\cite{Pablo_PhD} & PS + 1.5 g/l PPO & $\sigma_\text{int}=(1.69\pm0.23)$& $1.7\pm0.5$ & $1.3\pm0.3$\tabularnewline
  &    &   @ 156 keV &   & \tabularnewline
\hline 
\end{tabular}
\caption{The universal parameter $v_1^\text{int}$ and the corresponding contribution to the energy resolution at 1 MeV, $R_\text{int}(\text{1 MeV}$), for organic scintillators considered in this paper. We add also two measurements from Borexino detector, considered in~\cite{MyIntrinsic}.}
\label{Table:AllIRData}
\end{table*}

\section{Recommendations for future measurements}

Based on the data analysis presented in this paper, we propose the following recommendations for future precision measurements:

\begin{enumerate}
\item 
Report energy resolution in photoelectron (p.e.) units rather than energy scales. This eliminates uncertainties associated with nonlinear energy-to-charge conversion and avoids the need to convert energy back to p.e.
\item Maximize light yield in the experimental setup to improve measurement precision.
\item Use a multi-PMT configuration (2 or 4 PMTs), as demonstrated in~\cite{BC408ep}. This approach:
\begin{itemize}
    \item reduces systematic uncertainties from statistical term subtraction;
    \item simplifies the evaluation of single-electron response (SER) parameter uncertainties;
    \item further decreases PMT non-uniformity contributions ($v(p)$) as more PMTs are added.
\end{itemize}
\item Prefer the Wide-Angle Compton Coincidence (WACC) technique over electron guns, as it generally yields better performance. However, non-monochromatic effects must still be accounted for at low energies.

\item Include measurements with monoenergetic protons, which enable precise determination of the $v(p)$ contribution—crucial for accurate high-energy analysis.
\end{enumerate}

\section{Acknowledgements}
  This work was supported by the Russian Science Foundation (RSF) (Grant
No. 24-12-00046).

\end{document}